\documentclass[aps,amsmath,amssymb,twocolumn,showpacs,prl]{revtex4}
\usepackage{graphicx,color}
 \graphicspath{{./}{./figures/}}
\usepackage{dcolumn}
\usepackage{bm}
\newcommand{\red}{\color{red}}
\newcommand{\black}{\color{black}}

\newcommand{\be}{\begin{equation}}
\newcommand{\ee}{\end{equation}}
\newcommand{\bea}{\begin{eqnarray}}
\newcommand{\eea}{\end{eqnarray}}

\newcommand{\nn}{\nonumber }

\newcommand{\rme}{\mathrm{e}}
\newcommand{\rmd}{\mathrm{d}}

\newcommand{\fig}[2]{\includegraphics[width=#1]{./#2}}
\newcommand{\Fig}[1]{\includegraphics[width=8.7cm]{#1}}

\begin{document}

\title{Distribution of velocities in an avalanche}
\author{Pierre Le Doussal and Kay J\"org Wiese} \affiliation{CNRS-Laboratoire
de Physique Th{\'e}orique de l'Ecole Normale Sup{\'e}rieure, 24 rue
Lhomond,75005 Paris, France} 

\pacs{68.35.Rh}

\begin{abstract}For a driven elastic object near depinning, 
we  derive from first principles the distribution of instantaneous velocities in an avalanche. We prove that above the upper critical dimension, $d \geq d_{\mathrm{uc}}$, the $n$-times distribution of the center-of-mass velocity is equivalent to the 
prediction from the  ABBM stochastic equation. 
Our method allows to compute space and time dependence from an instanton equation. We extend the calculation beyond mean field, to
lowest order in $\epsilon=d_{\mathrm{uc}}-d$. 
\end{abstract}
\maketitle

Obtaining a quantitative description of the dynamics during an avalanche is
of great importance for systems whose dynamics is governed by jumps, 
such as magnets, superconductors, earthquakes, the contact line of fluids, or fracture
\cite{DSFisher1998, fracture,
LeDoussalWieseMoulinetRolley2009,mwalls,AltshulerJohansenPaltielJinBasslerRamosChenReiterZeldov2004
}.
In particular the motion of domain walls (DW) in magnets is important for many applications,
such as magnetic recording.
It can be measured from the  Barkhausen (magnetization) noise \cite{barkhausen,crackling}, which is a complicated time-dependent signal. Its origin
is due to an interplay between quenched impurities and the elastic deformation energy
which tend to pin the DW, as well as the driving and magnetostatic forces. 

A major step forward was accomplished by Alessandro,  Beatrice, Bertotti and Montorsi (ABBM) \cite{ABBM} who introduced, on a phenomenological basis, 
a stochastic equation
approximating the DW motion by a single degree of freedom. Although a crude
description, this model has been used extensively to compare with experiments on magnets,
with success in some ``mean-field like'' cases, and failure in other \cite{DurinZapperi2000}. However, 
no microscopic foundation for the validity of this model exists.

On the other hand, sophisticated field theoretic methods  were developed
in the last decades to study systems with quenched disorder. In particular, for elastic interfaces, relevant to
describe DW motion, functional RG methods (FRG)
\cite{Nattermann,Narayan-Fisher, DSFisher1998,us-dep} have recently allowed to derive
the distribution of quasi-static avalanche sizes \cite{LeDoussalMiddletonWiese2008,avalanchesFRG}. Until now however, no description of the dynamics
during an avalanche was available. In fact, since it involves much faster motion 
than the average driving velocity, it led to difficulties in the early FRG approaches
\cite{Narayan-Fisher}.

The aim of this Letter is to show how to compute from first principles the
distribution of instantaneous velocities in an avalanche. We study a single elastic
interface, of internal dimension $d$ (total space dimension is $D=d+1$) 
at zero temperature, near the depinning threshold. The method works in an expansion 
around the upper critical dimension $d_{\mathrm{uc}}$, with $d_{\mathrm{uc}}=4$ for standard elasticity, 
and $d_{\mathrm{uc}}=2$ in presence of long-range elasticity, e.g.\ arising from dipolar forces. Remarkably, we find that for $d = d_{\mathrm{uc}}$ (and above)
and in the scaling limit, the $n$-time probability distribution (with $n$ arbitrary) of the center 
of mass of the interface is  equivalent to that of the 
ABBM stochastic equation, in terms of renormalized parameters which in some
cases can be estimated. The two methods are rather different in spirit, and the
identification non-trivial. Our result establishes the universality of the ABBM model for $d \geq d_{\mathrm{uc}}$. In addition it allows to resolve the spatial structure, and gives the corrections to ABBM for $d<d_{\mathrm{uc}}$. 

Here we sketch a very simple derivation, for details and  {\em various subtleties} involved
we refer to \cite{us}. Consider the equation of motion, in the comoving frame, for the local velocity of
an interface driven at velocity $v$:
\begin{equation}\label{eqmo}
(\eta_0 \partial_t - \nabla^2_x ) \dot u_{xt} = \partial_t F(v t +
u_{xt} , x) - m^2 \dot u_{xt} \ . 
\end{equation}
It is obtained by time derivation (noted indifferently $\dot u$ or $\partial_t u$) of the standard overdamped equation of motion. Here $x$ is the $d$-dimensional
internal coordinate, $v t + u_{xt}$ the space and time dependent
displacement field and $\eta_0$ the friction. $F(u,x)$ is the quenched
random pinning force from the impurities, with  e.g.\ Gaussian distribution and variance $\overline{F(u,x) F(u',x')} = \delta^d(x-x') \Delta_0(u-u')$. $m^2$ is the strength of the restoring force $-m^2 (u_{xt}-vt)$ (i.e.\  the mass, or spring constant), which flattens the interface beyond a scale $L_m\sim 1/m$. In the small $m$, large $L_m$, limit, studied here, the interface
has the roughness exponent $\zeta$ of the depinning transition, with $u \sim x^\zeta$ for $x \lesssim L_m$
and $u \sim L_m^{\zeta}$ for $L > L_m$. For simplicity we chose standard elasticity $\sim \nabla_x^2$, but it can be replaced by an arbitrary elastic kernel as needed in applications \cite{DurinZapperi2000,fracture,LeDoussalWieseMoulinetRolley2009}.

\begin{figure}
\Fig{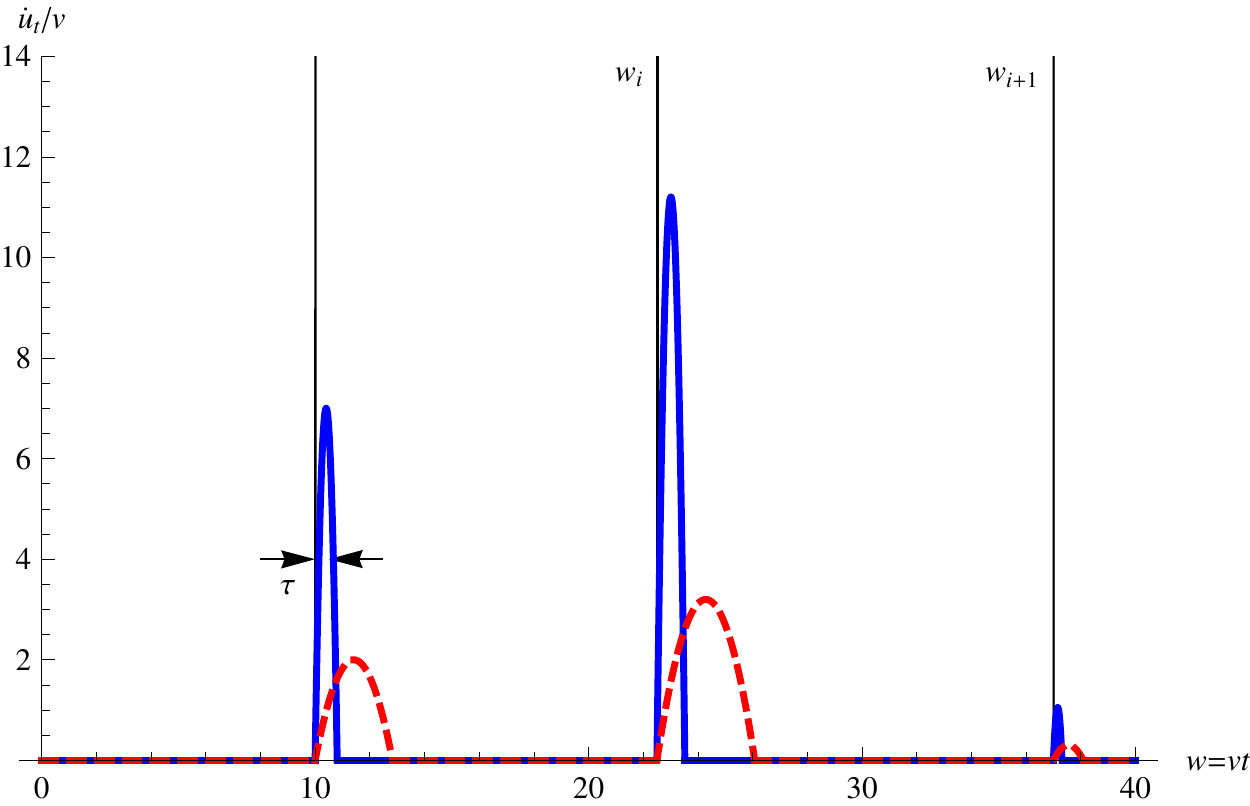}
\caption{Schematic plot of the instantaneous velocity (divided by $v$) as a function of $v t$ for different $v$. The area under the curve is
the avalanche size hence is constant as $v \to 0^+$. The quasi-static avalanche positions $w_i$ are indicated.}
\label{f:1}
\end{figure}

Near the depinning transition, i.e.\ at small $v$, 
the interface proceeds via avalanches. This is easiest seen 
in the center-of-mass position $u_t = L^{-d} \int_x u_{xt}$. There is a well-defined
quasi-static limit $v=0^+$ where $u_{t} = u(w)$, with $w=vt$ the well position. The  process $u(w)$  jumps
at discrete locations $w_i$, i.e.\  $u(w)=L^{-d} \sum_i S_i \theta(w-w_i)$, with
$S_i$  the avalanche sizes. Their statistics was predicted via FRG, and checked numerically \cite{LeDoussalMiddletonWiese2008,RossoLeDoussalWiese2009a, avalanchesFRG}. There, the bare disorder correlator $\Delta_0(u)$ flows, under coarse graining, to the renormalized one $\Delta(u)$, which, at the depinning transition exhibits a linear cusp $- \Delta'(0^+) >0$. This cusp is directly related to the moments of the
normalized size distribution $P(S)$, via \cite{avalanchesFRG}
\begin{equation} \label{rel1}
S_m := \frac{\langle S^2 \rangle}{2 \langle S \rangle} = \frac{|\Delta'(0^+)|}{m^4} \ .
\end{equation}
$S_m \sim m^{-(d+\zeta)}$ is the large-scale cutoff of $P(S)$. 
Here we study the dynamics {\it inside} these avalanches, which occur for  small $v$  on a time scale $\tau_m \sim L_m^{z} \ll \Delta w/v$, where $\Delta w$ is the typical separation of avalanches in the same space region, and $z$ the dynamical exponent. Hence we are considering small enough $v$ so that avalanches remain well separated, a condition equivalent to $L_m \ll \xi_v$, where $\xi_v$ is the standard critical correlation length \cite{Narayan-Fisher,Nattermann} near depinning (for $m=0$). This is illustrated on figure \ref{f:1}.

The information about the dynamics in an avalanche is contained in the $n$-times cumulants $C_n = \overline{\dot u_{t_1} \ldots \dot u_{t_n}}^c$, $n\geq 2$ (with $ \overline{\dot u_t}=0$). In the limit $v \to 0^+$ the product $\dot u_{t_1} \ldots  \dot u_{t_n}$ vanishes
unless all times are inside an avalanche. 
The probability that exactly one avalanche occurs in a 
time interval $T< \Delta w/v$ is $\rho_0 v T$, with $\rho_0=L^{d}/\langle S \rangle$ the avalanche density per unit $w$.
$C_n$ is thus $O(v)$, rather than $O(v^n)$, the hallmark of a non-smooth motion. 
 In addition, $C_n$ obeys the sum rule 
$L^{n d} \int_{[-T/2,T/2]^n} \rmd t _1\ldots \rmd t _n\, \overline{\dot u_{t_1} \ldots  \dot u_{t_n}} = \rho_0 v T \langle S^n \rangle+ O(v^2)$. 
It can be computed perturbatively in the (renormalized) disorder. For $n=2$ and to lowest order one finds
\begin{eqnarray} \label{2point}
 \overline{\dot u_{t_1} \dot u_{t_2}}^c \label{d2}  =  -  L^{-d}  \Delta'(0^+)  \frac{v}{m^2 \eta} \rme^{- \frac{m^2}{\eta} |t_1-t_2| }
\end{eqnarray}
where here and below $\eta$ is the renormalized friction \cite{footnote1}.
Integrating over time, one recovers (\ref{rel1}).

To obtain all moments at once, as well as the velocity distribution, we now compute the generating function
\begin{equation}
 Z[\lambda] = L^{-d} \partial_v \overline{e^{\int_{xt} \lambda_{xt} (v + \dot u_{xt})}}\Big|_{v=0^+} \ .
\end{equation}
The average over disorder (and initial conditions) is obtained
from the dynamical action $S = S_0 + S_{\mathrm{dis}}$ of (\ref{eqmo}):
\begin{eqnarray}
 S_0 &=& \int_{xt} \tilde u_{xt} (\eta \partial_t - \nabla_x^2 + m^2) \dot u_{xt}  \label{msr} \\
 S_{\mathrm{dis}} &=& -\frac{1}{2} \int_{xtt'}  \tilde u_{xt}  \tilde u_{xt'}  \partial_t \partial_{t'} \Delta(v (t-t') + u_{xt}-u_{xt'} ) \qquad \label{msr2}
\end{eqnarray}
This yields 
\begin{equation}\label{Z}
 Z[\lambda] = L^{-d} \partial_v \int {\cal D} [\dot u] {\cal D} [\tilde u]\, e^{- S + \int_{xt} \lambda_{xt} (v + \dot u_{xt}) }\Big|_{v=0^+} 
\end{equation}
with $Z[0]=0$. We write
\begin{eqnarray}\nn
\lefteqn{\partial_t \partial_{t'} \Delta(v (t-t') + u_{xt}-u_{xt'} ) }\\
& =&\nn
(v + \dot u_{xt}) \partial_{t'} \Delta'(v (t-t') +u_{xt}-u_{xt'}) \\
& =& (v + \dot u_{xt})  \Delta'(0^+) \partial_{t'}  {\rm sgn}(t-t') + \ldots  
\end{eqnarray} 
where we have used that the interface is only moving forward (Middleton theorem \cite{middleton}). We can thus rewrite the disorder term as
$S=S^{\mathrm{tree}}_{\mathrm{dis}}+\ldots $, where 
\begin{equation} \label{tree0}
 S^{\mathrm{tree}}_{\mathrm{dis}} =    \Delta'(0^+)  \int_{xt}  \tilde u_{xt}  \tilde u_{xt}  (v + \dot u_{xt}) 
\end{equation}
is the so-called tree-level or mean-field action \cite{footnote1}. The terms neglected are $O(\Delta''(0^+))$ and higher derivatives, and
we have shown that they contribute only to $O(\epsilon)$ to $Z[\lambda]$, hence can be
neglected at tree level. 

We now study the tree approximation for $Z[\lambda]$, i.e.\ (\ref{Z}) with $S_{\mathrm{dis}}$ replaced by (\ref{tree0}).
Thus the highly non-linear action (\ref{msr2}) has been reduced to a much simpler cubic theory! Even more
remarkably, $\dot u_{xt}$ appears only linearly in (\ref{tree0}), and viewing $\dot u$  as a response field, 
the tree level theory is {\it equivalent to the following non-linear equation}:
\begin{equation} \label{mfnonlinear}
 (\eta \partial_t + \nabla_x^2 - m^2) \tilde u_{xt} - \Delta'(0^+) \tilde u_{xt}^2  + \lambda_{xt} = 0
\end{equation}
We denote $\tilde u^\lambda_{xt}$ the solution of this equation for a given source $\lambda_{xt}$. 
Performing the derivative w.r.t $v$ in (\ref{Z}) gives 
\begin{eqnarray}
  Z[\lambda] &=& L^{-d} \int_{xt} \lambda_{xt} - \Delta'(0^+)   (\tilde u_{xt}^\lambda)^2  \\
& =&  L^{-d}  \int_{xt} (- \eta \partial_t - \nabla_x^2 + m^2) \tilde u^\lambda_{xt} = m^2 L^{-d} \int_{xt}  \tilde u^\lambda_{xt} 
\nonumber 
\end{eqnarray}
where we have used equation (\ref{mfnonlinear}) and, in the last equality, assumed
that $\tilde u^\lambda$  vanishes  at large $t$ and $x$. 
To analyze the result,  it is convenient to 
use dimensionless equations, replacing $x \to x/m$,  $L \to L/m$,  $t \to \tau_m t$, $v \to v v_m$, $\lambda \to \lambda/S_m$ and $\tilde u_{xt} \to \tilde u_{xt}/m^2 S_m$, where $v_m=S_m m^{d}/\tau_m$, and $\tau_m = \eta/m^2$. From now on we
use these units, and consider the 
 center-of-mass velocity, thus  choosing $\lambda_{xt} = \lambda_t$ uniform. 
 
 The 1-time probability at time $t=0$ is  given by  $\lambda_t=\lambda \delta(t)$ through its Laplace transform
\begin{equation}\label{12}
\tilde Z(\lambda) = L^{-d} \partial_v \overline{e^{L^d \lambda (v + \dot u)}}\Big|_{v=0^+} \ .
\end{equation}
$\dot u=\dot u_{t=0}$ and the notation $\tilde Z$ reminds us that we use dimensionless units.
$\tilde u_{xt}=\tilde u_t$ and we need to solve
\begin{eqnarray} \label{mfnonlinear2}
(\partial_t - 1) \tilde u_{t} + \tilde u_{t}^2  = - \lambda \delta(t) 
\end{eqnarray}
with $\tilde u_{t} \to 0$ at $t=\pm \infty$:
\be 
 \tilde u_t  = \frac{\lambda}{ \lambda + (1-\lambda) e^{-t}} \theta(-t) 
\ee
Inserting into (\ref{12}) gives
\begin{eqnarray}
 \tilde Z(\lambda)  = \int_t \tilde u_t =  - \ln(1 -  \lambda) \ .
\end{eqnarray}
Calling $\tau_i$ the duration of the $i$-th avalanche out of $N$, and defining $\langle \tau \rangle:=\frac{1}{N} \sum_i \tau_i$
the mean duration,  the probability $p_{\mathrm{a}}$ that $t=0$ belongs to an avalanche is $p_ {\mathrm{a}} =
\rho_0 v \langle \tau \rangle$. Hence the total 1-time velocity probability 
is $P(\dot u) = (1-p_ {\mathrm{a}}) \delta(v+\dot u) + p_ {\mathrm{a}} \tilde P(\dot u)$
where $\tilde P(\dot u)$ is the probability given that $t=0$ belongs to an avalanche.
Both $\tilde P$ and $P$ are normalized to unity. One notes the two (always) exact relations
$\langle \dot u \rangle_P=0$, $p_ {\mathrm{a}} \langle \dot u +v \rangle_{\tilde P} = v$. Hence
for $v=0^+$ one has $\rho_0 \langle \tau \rangle \langle \dot u \rangle_{\tilde P}=1$
and, in dimension\-full units $Z(\lambda) = \frac{1}{m^d v_m} \tilde Z(m^d v_m \lambda) = 
L^{-d} \rho_0 \langle \tau \rangle \int \rmd \dot u\, \tilde P(\dot u) 
(e^{L^d \lambda \dot u}-1)$. We thus obtain, in the slow driving limit, the instantaneous velocity distribution in the range
$v_0 \ll \dot u \sim \tilde v_m$ ($v_0$ being a small velocity cutoff):
\begin{equation} \label{1time}
 \tilde P(\dot u) = \frac{1}{\rho_0 \langle \tau \rangle \tilde v_m^2}\, p\Big(\frac{\dot u}{\tilde v_m}\Big) \ , \qquad p(x) = \frac{1}{x} e^{-x}\ .
\end{equation}
We defined $\tilde v_m=(m L)^{-d} v_m =L^{-d} S_m/\tau_m$. Hence
$\langle \dot u \rangle_{\tilde P} \approx \tilde v_m/\ln(\frac{\tilde v_m}{v_0})$. Note that
(i) $p(x)$ is not a probability, but is normalized by $\int \rmd x\, x\, p(x)=1$ (ii) the
quantity which is distributed according to $p(x)$ is $x=\tau_m \int_x \dot u_{xt}/S_m$, which does not
contain the factor $L^{-d}$.

Similarly  one  obtains the $n$-time distribution of the center-of-mass velocity
solving (\ref{mfnonlinear2}) with $\lambda_t = \sum_{j=1}^n \lambda_j
\delta(t-t_j)$, noting $z_{ij}:=1-\rme^{-|t_{i}-t_{j}|/\tau_{m}}$
\begin{equation}\label{a98}
\tilde Z_{n} (\lambda_{1},\dots
,\lambda_{n}) = -\ln\! \left(\sum_{\Lambda \subset \{{1},\dots
,{n} \}  } \prod_{{i} \in \Lambda} [-\lambda_{i}]
\prod_{\{i,j \}\subset \Lambda , i<j}\! z_{ij}  \right)
\end{equation} For $n=2$
one finds $\tilde Z_{2} = -\ln (1-\lambda_{1}-\lambda_{2} +\lambda_{1}\lambda_{2} z)$ 
with $z=1-\rme^{- |t_2-t_1|/\tau_m}$. From this we obtain (i) the probability $q_{12}= v q'_{12}$
that both $t_1$ and $t_2$ belong to the same avalanche 
and the velocity distribution $\tilde P$ conditioned
to this event:
\begin{align}\label{2time}
& q'_{12} \tilde P(\dot u_1,\dot u_2) = \frac{1}{\tilde v_m^3} p\Big(\frac{\dot u_1}{\tilde v_m},\frac{\dot u_2}{\tilde v_m}\Big) \\
\label{p12}
&p (v_{1},v_{2})= 
\frac{e^{ - \frac{t}{2} -\frac{v_1 + v_2}{1 - e^{-t}}} }{(1-e^{-t})
\sqrt{v_1 v_2}}\,  {I_1\!\bigg(\frac{2\, e^{-t/2} \sqrt{v_1 v_2}}{1-e^{-t}}\bigg)}
\end{align}
with $t=|t_{2}-t_{1}|/\tau_{m}$,  $q'_{12} \tilde v_m=\ln(1/z)$, and $I_{1} (x)$ is the Bessel-$I$
function of the first kind. 
The probability that $t_1$  but not $t_2$  belongs 
to an avalanche is
\begin{equation} \label{2time2}
 q'_1 \tilde P_1(\dot u_1) =  \frac{1}{\tilde v_m^2} p\Big(\frac{\dot u_1}{\tilde v_m}\Big) \  , \qquad  p(\dot u_1) = \frac{e^{- \dot u_1/z}}{\dot u_1}
\end{equation} 
with $p'_a=q'_1+q'_{12}$. Since the probability that there exists an avalanche starting in $[t_1,t_1+dt_1]$ and ending in $[t_2,t_2+dt_2]$
is $- dt_1 dt_2 \partial_{t_1} \partial_{t_2} q_{12}$ we obtain the distribution of
durations $\tau$ as
\begin{eqnarray}
P(\tau) = \frac{1}{\rho_0 \tilde v_m \tau_m^2} \frac{e^{-\tau/\tau_m}}{(1-e^{-\tau/\tau_m})^2} \ .
\end{eqnarray}
For small durations $\tau \ll \tau_m$, 
 $P(\tau) \approx  \frac{1}{\rho_0 \tilde v_m \tau^2}$, cut off at $\tau \approx \tau_0$. 
This gives $\langle \tau
\rangle = \frac{1}{\rho_0 \tilde v_m} \ln(\frac{\tau_m}{\tau_0})$ in
good agreement with the above, using $\ln(\frac{\tau_m}{\tau_0}) \approx \ln(\frac{\tilde v_m}{v_0})$. Note that
$q'_{12} \tilde P(0^+,0^+)$ is proportional to the probability that an avalanche starts at $t_1$ and ends at $t_2$.
\begin{figure}{
\setlength{\unitlength}{0.87mm}
\fboxsep0mm
\red 
\mbox{\begin{picture} (100,52)\black
\put(14,0){\fig{7.1cm}{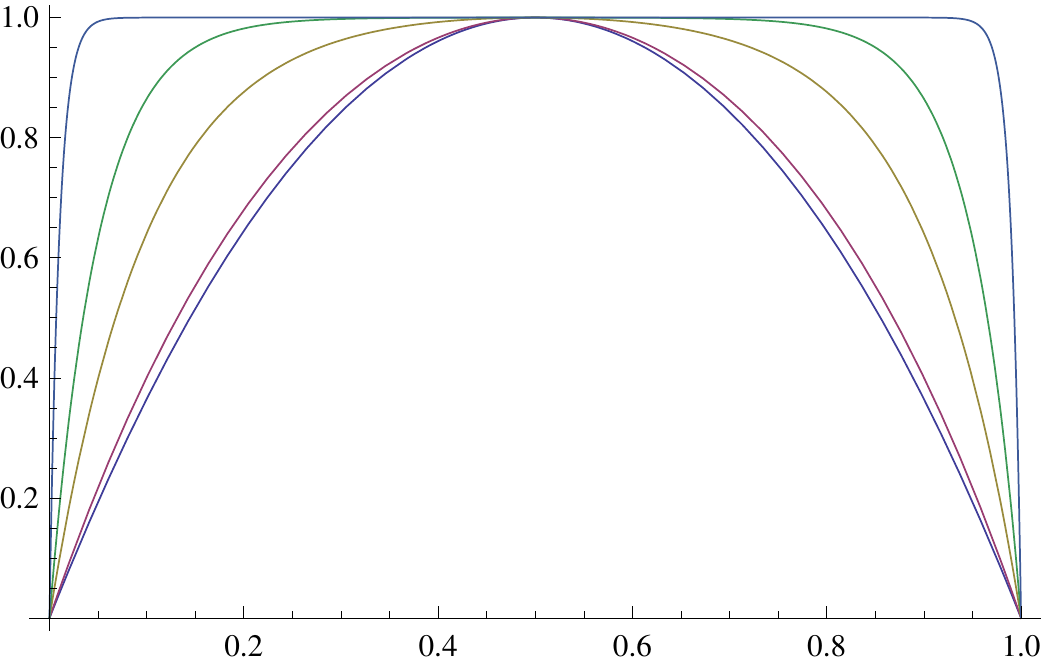}}
\put(96,2.8){$\displaystyle \frac{t}{\tau}$}
\put(0,46){$\displaystyle \frac{\overline{\dot{u}_{2} }}{\overline{\dot{u}_{2} }|_{t=\frac{\tau}2}}$} 
\end{picture}}}
\caption{``Pulse-shape'': The normalized velocity at time $t$ in an avalanche of
duration $\tau$ for $\tau \ll \tau_{m}$ (lower curve) to $\tau \gg \tau_{m}$ (upper
curve). }
\label{}
\end{figure}

The  ``shape'' of an avalanche with duration $\tau$ can then be extracted from the probabilities at 3 times $(t_1,t_2,t_3)=(0,t,\tau)$
setting $\dot u_1=\dot u_3=0^+$. From the generating function (\ref{a98})  for 3
times,  the probability distribution  for the intermediate-time
velocity is $P(\dot u_2)=b^{2} \dot
u_2 e^{-\dot u_2 b}$, with $\tilde v_m {b}:= \frac{1}{z_{12}}+\frac{1}{z_{23}}-1$
resulting in the average ``shape''
\begin{equation}\label{}
\overline{\dot u_2} = \frac{2}{b}= \tilde v_m \frac{4
\sinh\!\big(\frac{ t}{2\tau_{m}}\big) \sinh\!\big(\frac{
\tau}{2\tau_{m}} \big[1-\frac{t}{\tau}\big]\big)}{\sinh\!\big(\frac{
\tau}{2\tau_{m}} \big)}\ .
\end{equation}
This interpolates from a parabola for small $\tau \ll \tau_m$ to a flat
shape for the longest avalanches (see Fig 2.). This result holds for
an interface at or above its upper critical dimension, which previously
was used \cite{crackling} on the basis of the ABBM model. 

We now clarify the relation to the phenomenological  ABBM theory \cite{ABBM}. 
The latter models the interface as a single point driven
in a long-range correlated random-force landscape, $F(u)$, with {\it Brownian} statistics. It
amounts to suppressing the space dependence in (\ref{eqmo}), hence
corresponds in our  general model to the special case $d=0$ and
$\Delta_0(0)-\Delta_0(u)=\sigma |u|$. The instantaneous velocity ${\sf
v}=\dot u_t +v$ satisfies the stochastic  equation $\eta d {\sf
v} = m^2 (v - {\sf v} ) \rmd t\,  + \rmd F$ where $\overline{\rmd F^2}=2 \sigma {\sf v} \rmd t\, $, with associated Fokker-Planck equation
\begin{eqnarray} \label{fp}
\eta \partial_t Q = \partial_{{\sf v}} \left[\frac{\sigma}{\eta} \partial_{{\sf v}} ({\sf v} Q) + m^2 ({\sf v} - v) Q\right]
\end{eqnarray}
for the velocity probability $Q\equiv Q({\sf v},t|{\sf v}_1,0)$. For $v>0$ it evolves to the
stationary distribution $Q_0({\sf v}) = v_m^{-v/v_m} {\sf v}^{v/v_m-1} e^{- {\sf v}/v_m}/\Gamma(v/v_{m})$
with $v_m=S_m/\tau_m$ and here $S_m=\sigma/m^4$ and $\tau_m=\eta/m^2$. 
For $v=0^+$ one recovers (\ref{1time}), up to a normalization
which entails a small-scale cutoff. Similarly for $v=0^+$ one finds the propagator $Q({\sf v},t|{\sf v}_1,0)=v_m^{-1}
\tilde Q(\frac{{\sf v}}{v_m},\frac{t}{\tau_m}|\frac{{\sf v}_1}{v_m},0)$ with 
\begin{equation}\label{}
\tilde Q(v_2,t|v_1,0) = v_1 e^{v_1} \bigg[ p(v_1,v_2) +
\frac{1}{v_1} e^{-\frac{v_1}{1-e^{-t}}} \delta(v_2) \bigg]\ ,
\end{equation} and $p (v_{1},v_{2})$  given in Eq.~(\ref{p12}).  $\tilde Q(v_2,t|v_1,0)$
is solution of (\ref{fp}) with $Q({\sf v}_2,0^+|{\sf
v}_1,0)=\delta({\sf v}_2 - {\sf v}_1)$. 
The piece   $\sim \delta(v_2)$ 
 corresponds to avalanches which have already terminated at time $t$,
 and  is necessary for $Q$ to conserve probability. The {\it joint distribution}
$\tilde Q(v_2,t|v_1,0) \frac{1}{v_1}e^{-v_1}$ reproduces the 1-time and 2-times probabilities given in Eqs.~(\ref{2time}) and  (\ref{2time2}),  up to a global normalization. 
More generally, since $\sf{v} (t)$ is a Markov-process, the $n$-time velocity probability obtained from (\ref{mfnonlinear}) is
$q'_{1p} \tilde P(\dot u_1,\ldots ,\dot u_n)=\frac{1}{\dot u_1} e^{-\dot u_1}\prod_{j=1}^{n-1} Q(\dot u_{j+1} t_{j+1}| \dot u_j t_j)$.

Several remarks are in order. The  first one is    specific to the
ABBM model:  Since
it is the zero-dimensional limit of (\ref{eqmo}), 
the dynamical-action method can  be applied. Hence we just found
that for the ABBM model at $v=0^+$ {\it the tree approximation is
exact}. In the field theory  it means that the effective action $\Gamma$
equals the bare action $S$, and there are no loop corrections. Hence
$\Delta'(u)=\Delta_0'(u)=- \sigma \, \mbox{sgn}(u)$ is  an exact FRG fixed point (with $\zeta=4-d$) as noted 
in \cite{avalanchesFRG}. Crucial for this remarkable property is
that the force landscape is a Brownian, and even in $d=0$, this is
not valid for 
any other, e.g.\ shorter ranged, force landscape. In that sense, the
model proposed by ABBM \cite{ABBM}, although 
 unnatural from a microscopic point of view, appears extremely judicious. 

Second, since a realistic interface in a short-ranged random
force is described for $d \geq d_{\mathrm{uc}}$ by the tree approximation, we
proved that the temporal correlations of its center-of-mass velocity
for $v\to 0$ are given by the ABBM model. Only two parameters
enter, $\eta$ and $S_m$, which in $d=4$ acquire a logarithmic
dependence on $m$ \cite{avalanchesFRG}. 

Third, it is not expected that $S=\Gamma$ extends to finite driving
velocity $v>0$; hence
whether the  phenomenology of the ABBM model  with an avalanche
exponent $\tau$ dependent on $v$
has anything to do with realistic interface motion remains an open question. 

Fourth, the present theory  allows to go {\it beyond} the ABBM model
in several ways:  In $d \geq 4$, the non-linear equation
(\ref{mfnonlinear}) allows to study the full time- and
space-dependence of velocity correlations, as was done for the statics
in \cite{avalanchesFRG}.
Second, including loop-corrections  allows to compute corrections in a 
systematic expansion in $d=4 - \epsilon$ \cite{us}. 
The main result for the 1-time velocity distribution for $v_0 \ll v
\ll \tilde v_m$ is  to first order in $\epsilon$
\begin{equation}
P(v) \sim 1/v^{a} \  , \qquad a = 1 - \epsilon (1-\zeta_1)/3 + O(\epsilon^2)  
\end{equation}
i.e.\ $a=1-\frac{2}{9} \epsilon$ for a non periodic interface, and $a=1-\frac{\epsilon}{3}$ for
a charge density wave (CDW). The large-cutoff scale is
given by $\tilde v_m$ with $\eta_m \sim m^{2-z}$,  $z=2 - \frac{2}{9}
\epsilon$ for non-periodic disorder and $z=2-\frac{\epsilon}{3}$ for CDW \cite{Nattermann,Narayan-Fisher,us-dep}.

%
To conclude, we introduced a general method to compute both spatial and temporal
velocity correlations in an avalanche. Its tree-approximation is exact at and above the
upper critical dimension $d\ge d_{\mathrm{uc}}$. There the  center-of-mass motion is equivalent to
the phenomenological  ABBM
model. This establishes the range of validity of the latter. For $d < d_{\mathrm{uc}}$
corrections are calculated in a 
controled expansion in $\epsilon=d_{\mathrm{uc}}-d$. 

This work was supported by ANR grant 09-BLAN-0097-01/2 and in part
by  NSF grant PHY05-51164. We thank A.~Kolton and A.~Rosso for helpful discusions, and the 
KITP for hospitality.

\vfill 


\begin{thebibliography}{999}
\bibitem{footnote1}
We use 
the improved action, where  $\eta_0 \to \eta$ and $\Delta_0 \to \Delta$ in order to
obtain the correct result for $d \geq d_{\mathrm{uc}}$, see \cite{avalanchesFRG}. Note that $m^2$ and $v$ are
not corrected.
\end{thebibliography}
\end{document}